# Linguistic Uncertainty and Reply Engagement on X: A Cross-Domain Replication of the Uncertainty–Reply Asymmetry

**Mohamed Soufan**
Independent Researcher & Data Scientist
Antalya, Turkey
ORCID: https://orcid.org/0009-0004-1705-5574
Email: m@soufan.tech

## Abstract

Linguistic uncertainty is common in social media, but its relationship with engagement remains unclear across languages and topics. Using 2,258 English-language posts on Federal Reserve policy, inflation, and electoral politics collected over three days in April 2026, we test whether the Uncertainty–Reply Asymmetry observed in prior Arabic-language research replicates in a broader context. Posts are classified using a lexicon-based uncertainty framework, with approximately one-third identified as uncertain.

Uncertain posts receive 82% more replies on average than certain posts, with smaller increases in reposts and likes, replicating the asymmetric engagement pattern observed in prior work. Regression results confirm a positive and statistically significant association between uncertainty and replies ($\beta = 0.126$, $p = 0.011$), equivalent to ~13% higher expected reply engagement, while total engagement shows a positive but weaker association.

These findings suggest that linguistic uncertainty systematically increases conversational engagement and may reflect a general interactional mechanism across languages and domains.

## Introduction

Social media platforms have become central spaces for public discourse, where linguistic choices shape how information circulates and how audiences respond. Prior research has extensively examined the role of content features such as sentiment, emotional intensity, controversy, and misinformation in driving online engagement and shaping large-scale social and political behavior (Berger & Milkman, 2012; Bruns & Moe, 2014; Zhang & Counts, 2015).

However, relatively less attention has been paid to how linguistic uncertainty—expressions of doubt, incompleteness, or lack of verified information—affects not only the volume of engagement, but the form that engagement takes.

A recent study of Arabic-language discourse on X demonstrated that tweets expressing linguistic uncertainty were associated with significantly higher engagement than certainty-marked tweets, with the strongest association observed for replies (Soufan, 2026). This pattern, termed the Uncertainty–Reply Asymmetry, suggests that linguistically uncertain messages may function as interactional cues that invite conversational participation rather than passive endorsement. However, that study was conducted in a single language, on a single topic, in a context characterized by high political salience. Whether the effect reflects a general property of linguistic uncertainty in online communication, or is instead specific to Arabic-language discourse or context-specific conditions, remains an open question.

The present study addresses this question directly. We treat the Uncertainty–Reply Asymmetry as a testable hypothesis and subject it to a cross-linguistic, cross-domain replication using English-language posts on X. Specifically, we examine three topically distinct domains—Federal Reserve monetary policy, consumer price inflation, and electoral politics—chosen to represent contexts in which epistemic uncertainty is a natural and recurring feature of public discourse. We collect data from the same three-day window across all three topics and apply a lexicon-based uncertainty classifier adapted from the original Arabic framework. We then estimate regression models predicting engagement outcomes, with particular attention to whether the asymmetric effect on replies persists.

Our analysis indicates that uncertain posts receive more replies than certain posts across all three domains, and that this association remains positive and statistically significant in a model predicting log-transformed reply counts ($\beta = 0.126$, $p = 0.011$). The effect on total engagement is directionally consistent but approaches conventional significance thresholds in the pooled sample ($\beta = 0.151$, $p = 0.094$), likely reflecting the high variance inherent in cross-domain engagement data. Taken together, the results support the interpretation that the Uncertainty–Reply Asymmetry is unlikely to be an artifact of language or context, but reflects a broader interactional dynamic in which uncertainty signals openness and invites response.

This study contributes to the growing literature on computational social science and digital discourse in two respects. First, it provides cross-linguistic evidence for a named engagement effect, extending its empirical basis beyond Arabic to English and across multiple topical domains. Second, it demonstrates that a transparent, interpretable lexicon-based approach can recover consistent behavioral signals in large-scale observational data, even when classifier performance is conservative, reinforcing the value of replication as a methodological practice in computational social science.

## Related Work and Theoretical Framework

Research on social media engagement has identified a wide range of content and contextual factors associated with audience response, including emotional intensity, sentiment polarity, moral language, controversy, and misinformation. Across platforms, engagement is commonly operationalized as aggregated behavioral signals such as likes, shares, and replies, which are often treated as broadly comparable indicators of attention or popularity. While this literature has yielded important insights into what drives engagement volume, it has paid comparatively less attention to differences between engagement types and the interactional dynamics they represent.

A growing body of work distinguishes between passive and active forms of engagement. Likes and reposts primarily function as low-cost signals of approval or endorsement, whereas replies require greater communicative effort and typically reflect conversational involvement. This distinction suggests that engagement is not a unitary outcome but a heterogeneous set of behaviors with different social meanings. Despite this, most large-scale studies of social media behavior continue to treat engagement as a single aggregated measure, obscuring variation in how audiences respond to different types of content.

Linguistic uncertainty has been studied across disciplines, including linguistics, psychology, and communication research, where it is associated with hedging, epistemic modality, and expressions of incomplete or provisional knowledge. In online environments, uncertainty is often discussed in relation to misinformation, rumor diffusion, and credibility assessment (Kwon et al., 2016), and frequently framed as a liability—a signal of low information quality or weak speaker commitment. This framing, however, overlooks the potential interactional functions of uncertainty in conversational settings.

From an interactional perspective, expressions of uncertainty may function as invitations for participation. Questions, hedges, and rumor markers can signal openness and incompleteness, thereby lowering barriers for audience response. Rather than positioning the speaker as a final authority, uncertainty-marked messages may implicitly solicit clarification and confirmation. This suggests that uncertainty may shape not only whether users engage, but how they engage—a distinction with direct implications for understanding conversational dynamics in digital public spheres.

Building on this perspective, Soufan (2026) examined the relationship between linguistic uncertainty and engagement in Arabic-language discourse on X, using a dataset of 16,695 tweets about Lebanon collected over a 35-day period. That study introduced a lexicon-based uncertainty classifier comprising 60 markers across six categories and found that uncertain tweets were associated with 51.5% higher mean total engagement, with the largest relative difference observed for replies (+81.5%) compared to reposts (+65.2%) and likes (+47.0%). Regression analyses confirmed a positive and significant association between uncertainty and total engagement ($\beta = 0.221$, $p < 0.001$), and the asymmetric pattern across engagement types was interpreted as evidence that linguistic uncertainty functions as an interactional cue rather than a broadcast signal, with the strongest effect observed for conversational forms of engagement. This motivates the present study, which tests the generalizability of this pattern beyond a single language and context.

## Materials and Methods

## Study Design

This study employs a cross-domain replication design, applying the analytical framework developed in Soufan (2026) to English-language discourse on X across three topically distinct domains. The design holds constant the data collection period, classifier architecture, engagement operationalization, and statistical approach, while varying the language and topic context. This approach allows direct assessment of whether the Uncertainty–Reply Asymmetry identified in Arabic-language discourse generalizes to English.

## Data Collection

Data were collected from X using the Apify platform, an automated web-scraping tool that interfaces with the X web interface for data retrieval. Only publicly accessible posts were collected in accordance with platform terms of service. Three separate datasets were assembled, each corresponding to one topical domain: Federal Reserve monetary policy, consumer price inflation, and electoral politics. Each dataset was collected using a domain-specific boolean search query combining relevant topic terms with date constraints and operator-level filters excluding reposts and replies (-is:retweet -is:reply). The language filter was set to English for all three queries.

All three datasets were collected from an identical three-day window, spanning April 8 to April 10, 2026. This window was selected to ensure comparable exposure time across posts, sufficient for meaningful engagement accumulation while avoiding the high variance associated with very fresh content. The selected window excludes very recent posts, for which engagement counts are systematically underestimated, as well as older posts for which accumulated engagement reflects prolonged exposure rather than immediate audience response.

Unlike the original Arabic study, replies were excluded at the collection stage rather than retained as part of the analytical sample. This design choice ensures that the dataset consists exclusively of original posts, and that reply counts reflect audience responses to those posts rather than being both predictors and outcomes in the same analysis. Quote tweets were retained where present.

Raw records totaled 1,494 for the Federal Reserve topic, 979 for CPI inflation, and 705 for US politics, yielding a combined initial sample of 3,178 posts. During preprocessing, records were filtered to retain only English-language posts (lang = en), deduplicated first by post identifier and then by exact text match to remove bot-generated repeated content, and filtered to exclude posts containing fewer than five words. A post-level topic relevance filter was additionally applied to each dataset, retaining only posts containing at least one domain-relevant keyword, in order to exclude posts that matched the search query incidentally through unrelated use of query terms. The final analytical sample comprised 629 posts for the Federal Reserve topic, 934 for CPI inflation, and 695 for US politics, for a total of 2,258 posts across all three domains.

## Engagement Measures

Engagement was operationalized as the sum of likes, reposts, and replies received by each post. This composite measure reflects overall audience response and allows for the examination of individual engagement components in descriptive analyses. Given the highly skewed distribution of engagement values, the dependent variable in regression analyses was specified as the natural logarithm of one plus total engagement, log(1 + Total Engagement). A secondary regression model used log(1 + Reply Count) as the dependent variable, motivated by the theoretical centrality of replies as a conversational engagement form in the Uncertainty–Reply Asymmetry framework.

Linguistic Uncertainty Classification

Linguistic uncertainty was identified using a rule-based lexicon classifier adapted from the Arabic-language framework introduced in Soufan (2026). The English classifier comprises markers organized into six categories corresponding to the original Arabic taxonomy: modal verbs expressing epistemic possibility (e.g., *may*, *might*, *could*); hedges signaling speaker uncertainty (e.g., *I think*, *seems*, *probably*, *apparently*); question markers indicating information-seeking (e.g., interrogative constructions, bare question marks); contextual who-questions identifying agent or source uncertainty (e.g., *who is responsible*, *who's behind*); information uncertainty phrases expressing explicit doubt (e.g., *not sure*, *unclear*, *remains to be seen*); and rumor markers signaling unverified information (e.g., *reportedly*, *allegedly*, *sources say*, *speculation*). The full lexicon is provided as Supplementary Material.

Posts were classified as uncertain if they contained one or more markers from any category. Context-sensitive rules were applied to reduce false positives; for instance, question marks were treated as uncertainty markers only in interrogative constructions, and modal verbs were matched on word boundaries to avoid partial matches.

No formal validation study against human annotation was conducted for the English classifier, representing a deviation from the original Arabic study, which achieved an overall accuracy of 73.5% and a recall of 1.00 against native speaker annotation. The present classifier is expected to behave conservatively, likely underestimating the true prevalence of uncertainty through false negatives rather than inflating it through false positives. As in the original study, this pattern of errors implies that any measurement error attenuates observed differences between uncertain and certain posts, suggesting that reported effect sizes represent conservative estimates. An attempt to expand the lexicon with additional financial expectation terms increased classification noise and attenuated estimated effects, consistent with a precision–recall tradeoff; the baseline lexicon was retained for all reported analyses.

Using the baseline classifier, 278 posts (44.2%) were classified as uncertain in the Federal Reserve dataset, 274 (29.3%) in the CPI inflation dataset, and 205 (29.5%) in the US politics dataset, for an overall rate of 757 uncertain posts (33.5%) in the pooled sample.

Control Variables

Regression models included two control variables. Post length was measured as the number of characters in the post text. URL presence was coded as a binary indicator of whether the post text contained a hyperlink, derived by detecting the string *http* in the post body. Account verification status, included as a control in the original Arabic study, was unavailable in the present dataset due to limitations of the data collection instrument and was therefore excluded from all models. Standard errors were clustered at the author level using the posting account username, extracted from the post URL, as a proxy for author identity, in order to account for potential non-independence arising from multiple posts by the same account.

### Statistical Analysis

The primary analysis employed ordinary least squares regression to estimate the association between linguistic uncertainty and reply engagement. The model was specified as:

$$\log(1 + \text{Reply Count}) = \beta_0 + \beta_1(\text{Uncertainty}) + \beta_2(\text{Post Length}) + \beta_3(\text{Has URL}) + \varepsilon$$

A secondary model replaced the dependent variable with log-transformed total engagement. For the pooled cross-domain analysis, topic fixed effects were included as binary indicators for two of the three domains, with CPI inflation serving as the reference category. Standard errors were clustered at the author level in all models. Statistical significance was assessed using two-tailed tests.

As robustness checks, models were re-estimated following winsorization of the top one percent of engagement values, trimming of the top one percent of observations, and specification of the pooled total engagement model using negative binomial regression to account for overdispersion in engagement count data. All analyses were conducted in Python 3.11 using the pandas, numpy, scipy, and statsmodels libraries.

### Ethical Considerations

This study analyzed publicly available posts collected from X. No private accounts, protected content, or personal communications were accessed. All analyses are reported at the aggregate level, and no individual post text, usernames, or identifying information are disclosed. The study complies with X's Terms of Service for academic research and adheres to established ethical guidelines for the analysis of public social media data. Given the public nature of the data and the absence of personal identification, the study poses minimal risk to users.

## Results

### Descriptive Engagement Differences

We first examine descriptive differences in engagement between posts expressing linguistic uncertainty and those expressing certainty, separately for each topic domain and in the pooled sample.

Figure 1. Reply Engagement by Topic and Uncertainty Condition

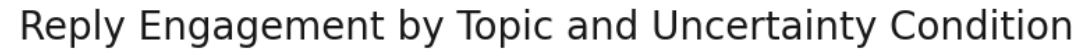

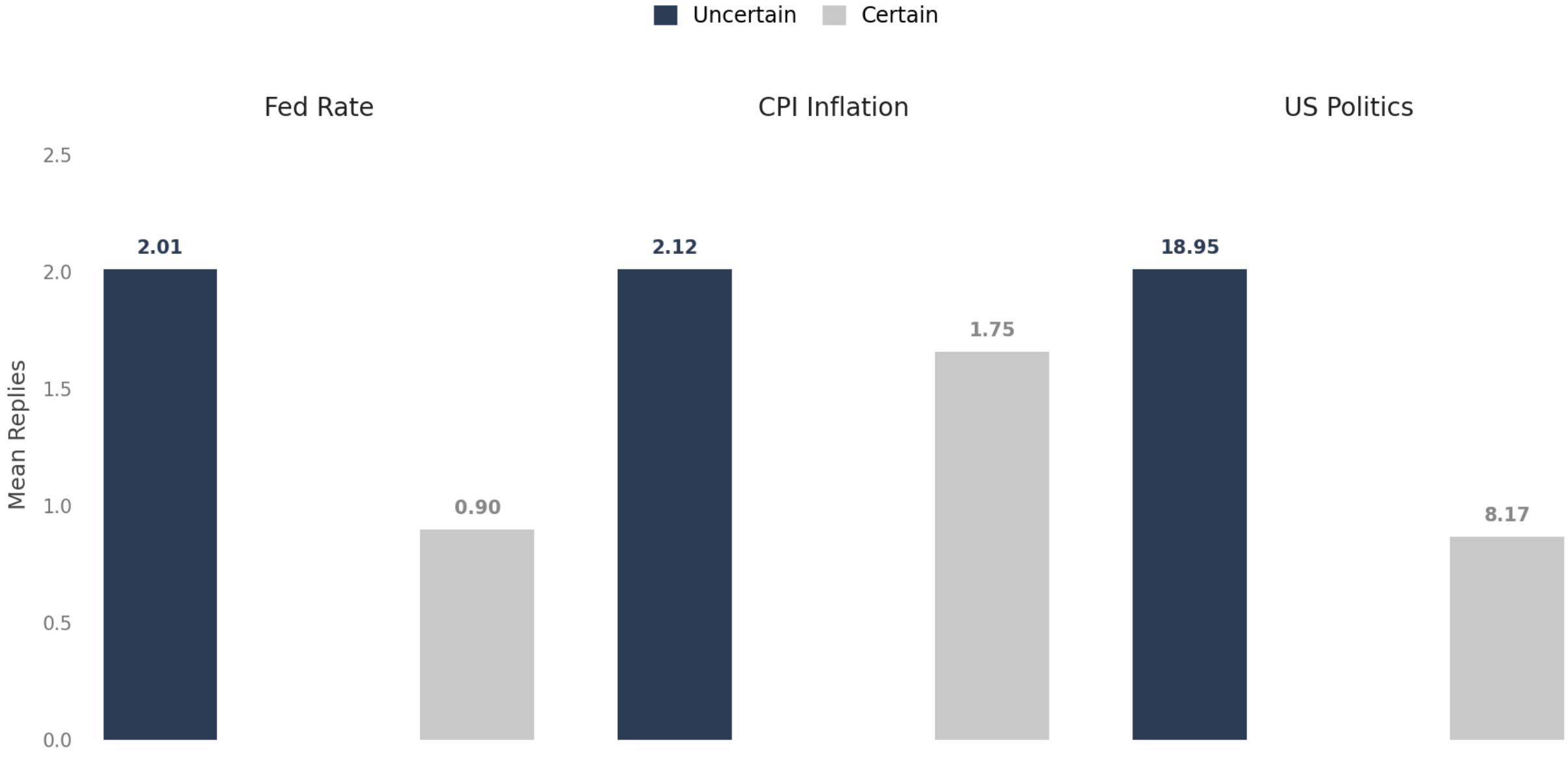

Across all three topics, uncertain posts receive more replies than certain posts.

Table 1. Descriptive Statistics by Topic

| Topic | N | % Uncertain | Mean Replies | | Mean Likes | | Mean Reposts | |
|---|---|---|---|---|---|---|---|---|
| | | | Uncertain | Certain | Uncertain | Certain | Uncertain | Certain |
| Fed Rate | 629 | 44.2% | 2.01 | 0.90 | 27.86 | 14.21 | 6.88 | 5.30 |
| CPI Inflation | 934 | 29.3% | 2.12 | 1.75 | 19.59 | 16.29 | 3.38 | 2.08 |
| US Politics | 695 | 29.5% | 18.95 | 8.17 | 175.60 | 160.03 | 35.71 | 29.65 |
| **Pooled** | **2,258** | **33.5%** | **6.64** | **3.65** | **64.87** | **62.73** | **13.42** | **11.83** |

***Note.*** Means are raw (untransformed) values. Regression models use log(1 + engagement count) as the dependent variable. Posts classified as uncertain contain at least one lexicon marker (modal verb, hedge, question mark, uncertainty phrase, or rumor marker).

In the Federal Reserve dataset (N = 629; 278 uncertain, 351 certain), uncertain posts received higher engagement across all three components. Compared to certain posts, uncertain posts received 122.2% more replies (M = 2.01 vs. 0.90), 96.0% more likes (M = 27.86 vs. 14.21), and 29.7% more reposts (M = 6.88 vs. 5.30). The largest relative difference was observed for replies.

In the CPI inflation dataset (N = 934; 274 uncertain, 660 certain), uncertain posts received more replies than certain posts (M = 2.12 vs. 1.75, +21.6%), along with more reposts (M = 3.38 vs. 2.08, +62.2%) and likes (M = 19.59 vs. 16.29, +20.2%).

In the US politics dataset (N = 695; 205 uncertain, 490 certain), the differences were largest. Uncertain posts received 132.0% more replies (M = 18.95 vs. 8.17), 20.4% more reposts (M = 35.71 vs. 29.65), and 9.7% more likes (M = 175.60 vs. 160.03). As in the other domains, the relative difference was largest for replies.

In the pooled sample (N = 2,258; 757 uncertain, 1,501 certain), uncertain posts received 82.0% more replies (M = 6.64 vs. 3.65), 13.4% more reposts (M = 13.42 vs. 11.83), and 3.4% more likes (M = 64.87 vs. 62.73).

Figure 2. Engagement by Uncertainty Condition

Uncertain posts receive higher engagement across replies, reposts, and likes, with the largest difference observed for replies.

Across all three domains and in the pooled sample, the largest relative difference between uncertain and certain posts was consistently observed for replies, with smaller differences for reposts and likes. This pattern is consistent with the Uncertainty–Reply Asymmetry described in the original Arabic-language study.

Primary Result: Replies Model

To assess whether the descriptive pattern persists after accounting for post-level covariates, we estimated an OLS regression model predicting log-transformed reply counts.

Table 2. OLS Regression Results: Uncertainty and Engagement

| | Model 1 *log(1 + Replies)* | Model 2 *log(1 + Replies)* Winsorized | Model 3 *log(1 + Total Engagement)* |
|---|---|---|---|
| **Key Variable** | | | |
| Uncertainty | **0.126**** (0.050) p = 0.011 | **0.119**** (0.048) p = 0.012 | **0.151**† (0.090) p = 0.094 |
| *95% CI* | [0.029, 0.224] | [0.026, 0.212] | [−0.026, 0.328] |
| **Controls** | | | |
| Post length | **0.0001*** (0.000) | **0.0001*** (0.000) | **0.0003*** (0.000) |
| URL present | −0.042 (0.044) | −0.041 (0.043) | −0.088 (0.087) |
| Topic FE: Fed Rate | **−0.188***** (0.041) | **−0.187***** (0.041) | **−0.538***** (0.082) |
| Topic FE: US Politics | **0.200**** (0.063) | **0.174**** (0.060) | 0.176 (0.114) |
| Constant | **0.358***** (0.043) | **0.359***** (0.042) | **1.146***** (0.091) |
| **Model Fit** | | | |
| $R^2$ | 0.029 | 0.028 | 0.032 |
| Adjusted $R^2$ | 0.027 | 0.026 | 0.030 |
| N | **2,258** | **2,258** | **2,258** |

***Note.*** OLS = ordinary least squares with author-level clustered standard errors (SE in parentheses). Dependent variables: log(1 + Reply Count) for Models 1–2; log(1 + total engagement) for Model 3. Model 2 uses winsorized reply counts at the 99th percentile (cap = 88 replies). Topic fixed effects use CPI Inflation as the reference category. Uncertainty is a binary indicator equal to 1 if the post contains at least one lexicon marker. † p < .10; * p < .05; ** p < .01; *** p < .001. Post length coefficient is small due to scaling (characters)

The model included linguistic uncertainty as the primary independent variable and controlled for post length and URL presence, with standard errors clustered at the author level.

The results indicate a positive and statistically significant association between linguistic uncertainty and reply engagement. Uncertain posts received significantly more replies than certain posts (β = 0.126, SE = 0.050, 95% CI [0.029, 0.224], p = 0.011). Given the log-transformed dependent variable, this coefficient corresponds to approximately 13% higher expected reply counts for uncertain posts, holding other variables constant. This effect remains statistically significant after winsorization of high-engagement outliers (β = 0.119, p = 0.012), confirming that the result is not driven by extreme values. Post length was positively associated with replies (β = 0.0001, p = 0.047). URL presence was not significantly associated with reply counts in this specification. The model explains a modest proportion of variance in reply engagement ($R^2$ = 0.029), consistent with prior research on the inherent unpredictability of social media attention dynamics.

Pooled Total Engagement

A secondary model replaced the dependent variable with log-transformed total engagement and included topic fixed effects for the Federal Reserve and US politics domains, with CPI inflation as the reference category. The uncertainty coefficient remained positive ($\beta = 0.151$, $SE = 0.090$, 95% CI [−0.025, 0.327], $p = 0.094$), corresponding to approximately 16% higher expected total engagement for uncertain posts. Post length was again positively associated with engagement ($\beta = 0.0003$, $p = 0.018$). The Federal Reserve topic was associated with significantly lower total engagement than the CPI reference category ($\beta = -0.538$, $p < 0.001$), reflecting differences in baseline engagement levels across domains. The model $R^2$ was 0.032.

### Robustness Checks

Results were robust across alternative specifications. Following winsorization of the top one percent of total engagement values (cap = 1,299), the uncertainty coefficient in the total engagement model was virtually unchanged ($\beta = 0.149$, $SE = 0.088$, $p = 0.089$). Trimming the top one percent of observations yielded similar results ($\beta = 0.133$, $SE = 0.081$, $p = 0.103$, $N =$ 2,235). The stability of the uncertainty coefficient across these specifications indicates that the observed association is not an artifact of high-engagement outliers.

A negative binomial regression model predicting raw total engagement counts was also estimated as a robustness check to account for overdispersion. The incidence rate ratio for uncertainty was 1.239 (95% CI [0.892, 1.721], $p = 0.201$), indicating a 24% higher expected engagement count for uncertain posts. Although this estimate does not reach conventional significance thresholds, the direction and magnitude are consistent with the OLS results. Convergence warnings during model fitting, likely attributable to extreme engagement outliers in the politics domain, suggest that these estimates should be interpreted with caution.

### Engagement Composition and the Uncertainty–Reply Asymmetry

Across all specifications, the association between linguistic uncertainty and engagement was not uniform across engagement types. In the pooled descriptive analysis, replies showed the largest relative difference between uncertain and certain posts (+82.0%), followed by reposts (+13.4%) and likes (+3.4%). This ordering—whereby conversational engagement (replies) shows a disproportionately stronger association with uncertainty than broadcast engagement (likes, reposts)—replicates the pattern observed in the original Arabic-language study and is consistent with the theoretical interpretation of linguistic uncertainty as an interactional cue.

The regression-based estimate of approximately 13% higher reply engagement for uncertain posts is smaller in magnitude than the raw descriptive difference of 82%, reflecting the contribution of correlated post-level features. However, the persistence of the association after adjustment strengthens the interpretation that linguistic uncertainty is associated with the observed difference in conversational engagement, rather than reflecting correlated content characteristics alone.

## Discussion

This study examined whether the Uncertainty–Reply Asymmetry—the tendency of linguistically uncertain posts to elicit disproportionately higher reply engagement relative to other forms of engagement—replicates in English-language discourse across multiple topical domains. The findings indicate that it does replicate. Across all three domains examined, uncertain posts received more replies than certain posts in descriptive comparisons, and this association remained positive and statistically significant in a regression model controlling for post length and URL presence ($\beta = 0.126$, $p = 0.011$). The effect on total engagement was directionally consistent across all topics, approaching conventional significance thresholds in the pooled model ($\beta = 0.151$, $p = 0.094$). These results extend the original Arabic-language findings to a new language and context, providing cross-linguistic evidence for the named effect.

The disproportionate association between uncertainty and replies, relative to likes and reposts, is theoretically meaningful. Likes and reposts function primarily as low-cost endorsement or dissemination signals, whereas replies require active communicative engagement and reflect conversational involvement. The consistent ordering observed across all three domains—replies showing the largest relative difference, followed by reposts and likes—suggests that linguistic uncertainty is more strongly associated with conversational participation than with passive or broadcast-oriented engagement. This pattern is consistent with the interpretation of uncertainty-marked messages as interactional cues that signal openness or incompleteness, thereby lowering the threshold for audience response. Rather than functioning as a deficit in information quality, linguistic uncertainty may implicitly invite clarification, speculation, or reaction from others, producing a shift in the composition of engagement toward more dialogic forms.

The regression-based estimate of approximately 13% higher reply engagement for uncertain posts is considerably smaller than the raw descriptive difference of 82%. This gap reflects the contribution of correlated post-level features, including post length and the presence of URLs, and is consistent with the pattern observed in the original Arabic study, where the adjusted regression estimate (approximately 25% higher total engagement) was similarly attenuated relative to the raw comparison (51.5% higher). The persistence of the association after adjustment, and its stability across winsorized and trimmed specifications, strengthens confidence that the observed pattern is not an artifact of outliers or confounded by basic content features.

It is also worth noting that the uncertainty classifier employed in this study is likely conservative, tending toward false negatives rather than false positives. This implies that posts classified as certain may include a proportion of genuinely uncertain content that the lexicon failed to detect. Under this condition, the true association between linguistic uncertainty and reply engagement may be larger than reported, and the present estimates should be understood as lower bounds rather than precise point estimates.

The cross-domain consistency of the effect is a central contribution of this study. The three domains examined—monetary policy, economic data, and electoral politics—differ substantially in subject matter, audience, and discourse conventions. The fact that the directional pattern holds across all three, despite these differences, suggests that the Uncertainty–Reply

Asymmetry may reflect a general feature of how linguistic uncertainty functions in online interaction rather than a property of any particular topic or community. At the same time, the magnitude of the effect varies across domains, with the largest descriptive differences observed in the US politics dataset and the smallest in the CPI inflation dataset. This variation may reflect differences in the salience of uncertainty within each domain, the degree to which audiences perceive uncertain claims as actionable or contestable, or differences in the engagement behavior of users who follow each type of content. These questions warrant further investigation.

The modest explanatory power of the regression models ($R^2 \approx 0.03$) is consistent with prior research on social media engagement, where a large proportion of variance in audience response is attributable to unobserved factors such as posting time, network position, follower count, and algorithmic exposure. Within this context, identifying a consistent and interpretable association between a linguistic feature and a specific form of engagement behavior represents a meaningful contribution, particularly given the cross-domain design and the conservative nature of the classifier.

## Conclusions

This study set out to examine whether the Uncertainty–Reply Asymmetry, originally identified in Arabic-language discourse on X, replicates in English-language posts across multiple topical domains. Using a dataset of 2,258 English-language posts spanning Federal Reserve monetary policy, consumer price inflation, and electoral politics, and applying a lexicon-based uncertainty classifier adapted from the original Arabic framework, we find consistent evidence that linguistically uncertain posts are associated with higher reply engagement than certain posts. A regression model predicting log-transformed reply counts confirms a statistically significant positive association ($\beta = 0.126$, $p = 0.011$), robust to winsorization of high-engagement outliers. The association with total engagement is directionally consistent and approaches conventional significance in the pooled model ($\beta = 0.151$, $p = 0.094$).

Across all three domains, the relative difference between uncertain and certain posts was largest for replies, smaller for reposts, and smallest for likes—replicating the asymmetric pattern that defines the Uncertainty–Reply Asymmetry. This ordering, consistent across domains and robust to alternative model specifications, supports the interpretation of linguistic uncertainty as an interactional cue that promotes conversational participation rather than passive endorsement.

By demonstrating that this pattern holds in English across three distinct topical contexts, the study provides cross-linguistic and cross-domain evidence for the generalizability of the Uncertainty–Reply Asymmetry. It also illustrates that transparent, interpretable lexicon-based classifiers can recover meaningful behavioral signals at scale, even under conservative classification conditions, and that replication across languages and domains is a productive methodological strategy for establishing robustness in computational social science research.

Future work could extend these findings by examining the Uncertainty–Reply Asymmetry in additional languages and platform contexts, developing validated uncertainty classifiers through human annotation, and investigating the qualitative content of replies to uncertain posts to assess whether they reflect clarification-seeking, contestation, or speculative engagement. Multilevel models that explicitly account for user-level heterogeneity would also allow more precise estimation of how individual-level factors moderate the relationship between linguistic uncertainty and engagement behavior.